\pdfoutput=1
\documentclass[conference]{IEEEtran}

\usepackage[acronyms,nonumberlist,nopostdot,nomain,nogroupskip,acronymlists={hidden}]{glossaries}
\usepackage{color}
\usepackage{amsmath}
\usepackage{caption}
\usepackage{subcaption}

\usepackage{graphicx}

\ifCLASSINFOpdf
\else
\fi
\begin{document}
\newacronym{3gpp}{3GPP}{3rd Generation Partnership Project}
\newacronym{4g}{4G}{4th generation}
\newacronym{5g}{5G}{5th generation}
\newacronym{6g}{6G}{6th generation}
\newacronym{5gc}{5GC}{5G Core}
\newacronym{aau}{AAU}{Active Antenna Unit}
\newacronym{abf}{ABF}{Analog Beamforming}
\newacronym{dbf}{DBF}{Digital Beamforming}
\newacronym{adc}{ADC}{Analog to Digital Converter}
\newacronym{aerpaw}{AERPAW}{Aerial Experimentation and Research Platform for Advanced Wireless}
\newacronym{ai}{AI}{Artificial Intelligence}
\newacronym{aimd}{AIMD}{Additive Increase Multiplicative Decrease}
\newacronym{am}{AM}{Acknowledged Mode}
\newacronym{amc}{AMC}{Adaptive Modulation and Coding}
\newacronym{amf}{AMF}{Access and Mobility Management Function}
\newacronym{aops}{AOPS}{Adaptive Order Prediction Scheduling}
\newacronym{api}{API}{Application Programming Interface}
\newacronym{apn}{APN}{Access Point Name}
\newacronym{ap}{AP}{Application Protocol}
\newacronym{aqm}{AQM}{Active Queue Management}
\newacronym{ausf}{AUSF}{Authentication Server Function}
\newacronym{avc}{AVC}{Advanced Video Coding}
\newacronym{awgn}{AGWN}{Additive White Gaussian Noise}
\newacronym{balia}{BALIA}{Balanced Link Adaptation Algorithm}
\newacronym{bbu}{BBU}{Base Band Unit}
\newacronym{bdp}{BDP}{Bandwidth-Delay Product}
\newacronym{ber}{BER}{Bit Error Rate}
\newacronym{bf}{BF}{Beamforming}
\newacronym{bler}{BLER}{Block Error Rate}
\newacronym{brr}{BRR}{Bayesian Ridge Regressor}
\newacronym{bs}{BS}{Base Station}
\newacronym{bsr}{BSR}{Buffer Status Report}
\newacronym{bss}{BSS}{Business Support System}
\newacronym{ca}{CA}{Carrier Aggregation}
\newacronym{caas}{CaaS}{Connectivity-as-a-Service}
\newacronym{cav}{CAV}{Connected and Autonoums Vehicle}
\newacronym{cb}{CB}{Code Block}
\newacronym{cc}{CC}{Congestion Control}
\newacronym{ccid}{CCID}{Congestion Control ID}
\newacronym{cco}{CC}{Carrier Component}
\newacronym{cd}{CD}{Continuous Delivery}
\newacronym{cdd}{CDD}{Cyclic Delay Diversity}
\newacronym{cdf}{CDF}{Cumulative Distribution Function}
\newacronym{cdn}{CDN}{Content Distribution Network}
\newacronym{cli}{CLI}{Command-line Interface}
\newacronym{cn}{CN}{Core Network}
\newacronym{codel}{CoDel}{Controlled Delay Management}
\newacronym{comac}{COMAC}{Converged Multi-Access and Core}
\newacronym{cord}{CORD}{Central Office Re-architected as a Datacenter}
\newacronym{cornet}{CORNET}{COgnitive Radio NETwork}
\newacronym{cosmos}{COSMOS}{Cloud Enhanced Open Software Defined Mobile Wireless Testbed for City-Scale Deployment}
\newacronym{cots}{COTS}{Commercial Off-the-Shelf}
\newacronym{cp}{CP}{Control Plane}
\newacronym{cyp}{CP}{Cyclic Prefix}
\newacronym{up}{UP}{User Plane}
\newacronym{cpu}{CPU}{Central Processing Unit}
\newacronym{cqi}{CQI}{Channel Quality Information}
\newacronym{cr}{CR}{Cell Reconfiguration}
\newacronym{cran}{C-RAN}{Centralized RAN}
\newacronym{crs}{CRS}{Cell Reference Signal}
\newacronym{csi}{CSI}{Channel State Information}
\newacronym{csirs}{CSI-RS}{Channel State Information - Reference Signal}
\newacronym{cu}{CU}{Central Unit}
\newacronym{d2tcp}{D$^2$TCP}{Deadline-aware Data center TCP}
\newacronym{d3}{D$^3$}{Deadline-Driven Delivery}
\newacronym{dac}{DAC}{Digital to Analog Converter}
\newacronym{dag}{DAG}{Directed Acyclic Graph}
\newacronym{das}{DAS}{Distributed Antenna System}
\newacronym{dash}{DASH}{Dynamic Adaptive Streaming over HTTP}
\newacronym{dc}{DC}{Dual Connectivity}
\newacronym{dccp}{DCCP}{Datagram Congestion Control Protocol}
\newacronym{dce}{DCE}{Direct Code Execution}
\newacronym{dci}{DCI}{Downlink Control Information}
\newacronym{dctcp}{DCTCP}{Data Center TCP}
\newacronym{dl}{DL}{Downlink}
\newacronym{dmr}{DMR}{Deadline Miss Ratio}
\newacronym{dmrs}{DMRS}{DeModulation Reference Signal}
\newacronym{drlcc}{DRL-CC}{Deep Reinforcement Learning Congestion Control}
\newacronym{dsrc}{DSRC}
{dedicated short-range communications}
\newacronym{d2d}{D2D}{device-to-device}
\newacronym{drs}{DRS}{Discovery Reference Signal}
\newacronym{du}{DU}{Distributed Unit}
\newacronym{e2e}{E2E}{end-to-end}
\newacronym{earfcn}{EARFCN}{E-UTRA Absolute Radio Frequency Channel Number}
\newacronym{ecaas}{ECaaS}{Edge-Cloud-as-a-Service}
\newacronym{ecn}{ECN}{Explicit Congestion Notification}
\newacronym{edf}{EDF}{Earliest Deadline First}
\newacronym{embb}{eMBB}{Enhanced Mobile Broadband}
\newacronym{empower}{EMPOWER}{EMpowering transatlantic PlatfOrms for advanced WirEless Research}
\newacronym{enb}{eNB}{evolved Node Base}
\newacronym{endc}{EN-DC}{E-UTRAN-\gls{nr} \gls{dc}}
\newacronym{epc}{EPC}{Evolved Packet Core}
\newacronym{eps}{EPS}{Evolved Packet System}
\newacronym{es}{ES}{Edge Server}
\newacronym{etsi}{ETSI}{European Telecommunications Standards Institute}
\newacronym[firstplural=Estimated Times of Arrival (ETAs)]{eta}{ETA}{Estimated Time of Arrival}
\newacronym{eutran}{E-UTRAN}{Evolved Universal Terrestrial Access Network}
\newacronym{faas}{FaaS}{Function-as-a-Service}
\newacronym{fapi}{FAPI}{Functional Application Platform Interface}
\newacronym{fdd}{FDD}{Frequency Division Duplexing}
\newacronym{fdm}{FDM}{Frequency Division Multiplexing}
\newacronym{fdma}{FDMA}{Frequency Division Multiple Access}
\newacronym{fed4fire}{FED4FIRE+}{Federation 4 Future Internet Research and Experimentation Plus}
\newacronym{fir}{FIR}{Finite Impulse Response}
\newacronym{fit}{FIT}{Future \acrlong{iot}}
\newacronym{fpga}{FPGA}{Field Programmable Gate Array}
\newacronym{fr2}{FR2}{Frequency Range 2}
\newacronym{fr1}{FR1}{Frequency Range 1}
\newacronym{fs}{FS}{Fast Switching}
\newacronym{fscc}{FSCC}{Flow Sharing Congestion Control}
\newacronym{ftp}{FTP}{File Transfer Protocol}
\newacronym{fw}{FW}{Flow Window}
\newacronym{ge}{GE}{Gaussian Elimination}
\newacronym{gnb}{gNB}{Next Generation Node Base}
\newacronym{gop}{GOP}{Group of Pictures}
\newacronym{gpr}{GPR}{Gaussian Process Regressor}
\newacronym{gpu}{GPU}{Graphics Processing Unit}
\newacronym{gtp}{GTP}{GPRS Tunneling Protocol}
\newacronym{gtpc}{GTP-C}{GPRS Tunnelling Protocol Control Plane}
\newacronym{gtpu}{GTP-U}{GPRS Tunnelling Protocol User Plane}
\newacronym{gtpv2c}{GTPv2-C}{\gls{gtp} v2 - Control}
\newacronym{gw}{GW}{Gateway}
\newacronym{harq}{HARQ}{Hybrid Automatic Repeat reQuest}
\newacronym{hetnet}{HetNet}{Heterogeneous Network}
\newacronym{hh}{HH}{Hard Handover}
\newacronym{hol}{HOL}{Head-of-Line}
\newacronym{hqf}{HQF}{Highest-quality-first}
\newacronym{hss}{HSS}{Home Subscription Server}
\newacronym{http}{HTTP}{HyperText Transfer Protocol}
\newacronym{hbf}{HBF}{Hybrid Beamforming}
\newacronym{ia}{IA}{Initial Access}
\newacronym{iab}{IAB}{Integrated Access and Backhaul}
\newacronym{ic}{IC}{Incident Command}
\newacronym{ietf}{IETF}{Internet Engineering Task Force}
\newacronym{imsi}{IMSI}{International Mobile Subscriber Identity}
\newacronym{imt}{IMT}{International Mobile Telecommunication}
\newacronym{iot}{IoT}{Internet of Things}
\newacronym{ip}{IP}{Internet Protocol}
\newacronym{itu}{ITU}{International Telecommunication Union}
\newacronym{kpi}{KPI}{Key Performance Indicator}
\newacronym{kpm}{KPM}{Key Performance Measurement}
\newacronym{kvm}{KVM}{Kernel-based Virtual Machine}
\newacronym{los}{LoS}{Line of Sight}
\newacronym{lsm}{LSM}{Link-to-System Mapping}
\newacronym{lstm}{LSTM}{Long Short Term Memory}
\newacronym{lte}{LTE}{Long Term Evolution}
\newacronym{lxc}{LXC}{Linux Container}
\newacronym{m2m}{M2M}{Machine to Machine}
\newacronym{mac}{MAC}{Medium Access Control}
\newacronym{manet}{MANET}{Mobile Ad Hoc Network}
\newacronym{mano}{MANO}{Management and Orchestration}
\newacronym{mc}{MC}{Multi-Connectivity}
\newacronym{mcc}{MCC}{Mobile Cloud Computing}
\newacronym{mchem}{MCHEM}{Massive Channel Emulator}
\newacronym{mcs}{MCS}{Modulation and Coding Scheme}
\newacronym{mec2}{MEC}{Multi-access Edge Computing}
\newacronym{mec}{MEC}{Mobile Edge Computing}
\newacronym{mfc}{MFC}{Mobile Fog Computing}
\newacronym{mgen}{MGEN}{Multi-Generator}
\newacronym{mi}{MI}{Mutual Information}
\newacronym{mib}{MIB}{Master Information Block}
\newacronym{miesm}{MIESM}{Mutual Information Based Effective SINR}
\newacronym{mimo}{MIMO}{Multiple Input, Multiple Output}
\newacronym{ml}{ML}{Machine Learning}
\newacronym{mlr}{MLR}{Maximum-local-rate}
\newacronym[plural=\gls{mme}s,firstplural=Mobility Management Entities (MMEs)]{mme}{MME}{Mobility Management Entity}
\newacronym{mmtc}{mMTC}{Massive Machine-Type Communications}
\newacronym{mmwave}{mmWave}{millimeter wave}
\newacronym{mpdccp}{MP-DCCP}{Multipath Datagram Congestion Control Protocol}
\newacronym{mptcp}{MPTCP}{Multipath TCP}
\newacronym{mr}{MR}{Maximum Rate}
\newacronym{mrdc}{MR-DC}{Multi \gls{rat} \gls{dc}}
\newacronym{mse}{MSE}{Mean Square Error}
\newacronym{mss}{MSS}{Maximum Segment Size}
\newacronym{mt}{MT}{Mobile Termination}
\newacronym{mtd}{MTD}{Machine-Type Device}
\newacronym{mtu}{MTU}{Maximum Transmission Unit}
\newacronym{mumimo}{MU-MIMO}{Multi-user \gls{mimo}}
\newacronym{mvno}{MVNO}{Mobile Virtual Network Operator}
\newacronym{nalu}{NALU}{Network Abstraction Layer Unit}
\newacronym{nas}{NAS}{Network Attached Storage}
\newacronym{nat}{NAT}{Network Address Translation}
\newacronym{nbiot}{NB-IoT}{Narrow Band IoT}
\newacronym{nfv}{NFV}{Network Function Virtualization}
\newacronym{nfvi}{NFVI}{Network Function Virtualization Infrastructure}
\newacronym{ni}{NI}{Network Interfaces}
\newacronym{nic}{NIC}{Network Interface Card}
\newacronym{now}{NOW}{Non Overlapping Window}
\newacronym{nsm}{NSM}{Network Service Mesh}
\newacronym{nr}{NR}{New Radio}
\newacronym{nrf}{NRF}{Network Repository Function}
\newacronym{nsa}{NSA}{Non Stand Alone}
\newacronym{nse}{NSE}{Network Slicing Engine}
\newacronym{nssf}{NSSF}{Network Slice Selection Function}
\newacronym{oai}{OAI}{OpenAirInterface}
\newacronym{oaicn}{OAI-CN}{\gls{oai} \acrlong{cn}}
\newacronym{oairan}{OAI-RAN}{\acrlong{oai} \acrlong{ran}}
\newacronym{oam}{OAM}{Operations, Administration and Maintenance}
\newacronym{ofdm}{OFDM}{Orthogonal Frequency Division Multiplexing}
\newacronym{olia}{OLIA}{Opportunistic Linked Increase Algorithm}
\newacronym{omec}{OMEC}{Open Mobile Evolved Core}
\newacronym{onap}{ONAP}{Open Network Automation Platform}
\newacronym{onf}{ONF}{Open Networking Foundation}
\newacronym{onos}{ONOS}{Open Networking Operating System}
\newacronym{oom}{OOM}{\gls{onap} Operations Manager}
\newacronym{opnfv}{OPNFV}{Open Platform for \gls{nfv}}
\newacronym{oran}{O-RAN}{Open RAN}
\newacronym{orbit}{ORBIT}{Open-Access Research Testbed for Next-Generation Wireless Networks}
\newacronym{os}{OS}{Operating System}
\newacronym{oss}{OSS}{Operations Support System}
\newacronym{pa}{PA}{Position-aware}
\newacronym{pase}{PASE}{Prioritization, Arbitration, and Self-adjusting Endpoints}
\newacronym{pawr}{PAWR}{Platforms for Advanced Wireless Research}
\newacronym{pbch}{PBCH}{Physical Broadcast Channel}
\newacronym{pcef}{PCEF}{Policy and Charging Enforcement Function}
\newacronym{pcfich}{PCFICH}{Physical Control Format Indicator Channel}
\newacronym{pcrf}{PCRF}{Policy and Charging Rules Function}
\newacronym{pdcch}{PDCCH}{Physical Downlink Control Channel}
\newacronym{pdcp}{PDCP}{Packet Data Convergence Protocol}
\newacronym{pdsch}{PDSCH}{Physical Downlink Shared Channel}
\newacronym{pdu}{PDU}{Packet Data Unit}
\newacronym{pf}{PF}{Proportional Fair}
\newacronym{pgw}{PGW}{Packet Gateway}
\newacronym{phich}{PHICH}{Physical Hybrid ARQ Indicator Channel}
\newacronym{phy}{PHY}{Physical}
\newacronym{pmch}{PMCH}{Physical Multicast Channel}
\newacronym{pmi}{PMI}{Precoding Matrix Indicators}
\newacronym{powder}{POWDER}{Platform for Open Wireless Data-driven Experimental Research}
\newacronym{ppo}{PPO}{Proximal Policy Optimization}
\newacronym{ppp}{PPP}{Poisson Point Process}
\newacronym{prach}{PRACH}{Physical Random Access Channel}
\newacronym{prb}{PRB}{Physical Resource Block}
\newacronym{psnr}{PSNR}{Peak Signal to Noise Ratio}
\newacronym{pss}{PSS}{Primary Synchronization Signal}
\newacronym{pucch}{PUCCH}{Physical Uplink Control Channel}
\newacronym{pusch}{PUSCH}{Physical Uplink Shared Channel}
\newacronym{qam}{QAM}{Quadrature Amplitude Modulation}
\newacronym{qci}{QCI}{\gls{qos} Class Identifier}
\newacronym{qoe}{QoE}{Quality of Experience}
\newacronym{qos}{QoS}{Quality of Service}
\newacronym{quic}{QUIC}{Quick UDP Internet Connections}
\newacronym{ra}{RA}{Resouces Allocation}
\newacronym{rach}{RACH}{Random Access Channel}
\newacronym{ran}{RAN}{Radio Access Network}
\newacronym[firstplural=Radio Access Technologies (RATs)]{rat}{RAT}{Radio Access Technology}
\newacronym{rbg}{RBG}{Resource Block Group}
\newacronym{rb}{RB}{Resource Block}
\newacronym{rcn}{RCN}{Research Coordination Network}
\newacronym{rc}{RC}{RAN Control}
\newacronym{rec}{REC}{Radio Edge Cloud}
\newacronym{red}{RED}{Random Early Detection}
\newacronym{renew}{RENEW}{Reconfigurable Eco-system for Next-generation End-to-end Wireless}
\newacronym{rf}{RF}{Radio Frequency}
\newacronym{rfc}{RFC}{Request for Comments}
\newacronym{rfr}{RFR}{Random Forest Regressor}
\newacronym{ric}{RIC}{\gls{ran} Intelligent Controller}
\newacronym{rlc}{RLC}{Radio Link Control}
\newacronym{rlf}{RLF}{Radio Link Failure}
\newacronym{rlnc}{RLNC}{Random Linear Network Coding}
\newacronym{rmr}{RMR}{RIC Message Router}
\newacronym{rmse}{RMSE}{Root Mean Squared Error}
\newacronym{rnis}{RNIS}{Radio Network Information Service}
\newacronym{rr}{RR}{Round Robin}
\newacronym{rrc}{RRC}{Radio Resource Control}
\newacronym{rrm}{RRM}{Radio Resource Management}
\newacronym{rru}{RRU}{Remote Radio Unit}
\newacronym{rs}{RS}{Remote Server}
\newacronym{rsrp}{RSRP}{Reference Signal Received Power}
\newacronym{rsrq}{RSRQ}{Reference Signal Received Quality}
\newacronym{rss}{RSS}{Received Signal Strength}
\newacronym{rssi}{RSSI}{Received Signal Strength Indicator}
\newacronym{rtt}{RTT}{Round Trip Time}
\newacronym{ru}{RU}{Radio Unit}
\newacronym{rus}{RSU}{Road Side Unit}
\newacronym{rw}{RW}{Receive Window}
\newacronym{rx}{RX}{Receiver}
\newacronym{s1ap}{S1AP}{S1 Application Protocol}
\newacronym{sa}{SA}{standalone}
\newacronym{sack}{SACK}{Selective Acknowledgment}
\newacronym{sap}{SAP}{Service Access Point}
\newacronym{sbt}{SBT}{Scheduler-based throttling}
\newacronym{sc2}{SC2}{Spectrum Collaboration Challenge}
\newacronym{scef}{SCEF}{Service Capability Exposure Function}
\newacronym{sch}{SCH}{Secondary Cell Handover}
\newacronym{scoot}{SCOOT}{Split Cycle Offset Optimization Technique}
\newacronym{sctp}{SCTP}{Stream Control Transmission Protocol}
\newacronym{sdap}{SDAP}{Service Data Adaptation Protocol}
\newacronym{sdk}{SDK}{Software Development Kit}
\newacronym{sdm}{SDM}{Space Division Multiplexing}
\newacronym{sdma}{SDMA}{Spatial Division Multiple Access}
\newacronym{sdn}{SDN}{Software-defined Networking}
\newacronym{sdr}{SDR}{Software-defined Radio}
\newacronym{seba}{SEBA}{SDN-Enabled Broadband Access}
\newacronym{sgsn}{SGSN}{Serving GPRS Support Node}
\newacronym{sgw}{SGW}{Service Gateway}
\newacronym{si}{SI}{Study Item}
\newacronym{sib}{SIB}{Secondary Information Block}
\newacronym{sinr}{SINR}{Signal to Interference plus Noise Ratio}
\newacronym{sip}{SIP}{Session Initiation Protocol}
\newacronym{siso}{SISO}{Single Input, Single Output}
\newacronym{sla}{SLA}{Service Level Agreement}
\newacronym{sm}{SM}{Service Model}
\newacronym{smo}{SMO}{Service Management and Orchestration}
\newacronym{smsgmsc}{SMS-GMSC}{\gls{sms}-Gateway}
\newacronym{snr}{SNR}{Signal-to-Noise-Ratio}
\newacronym{son}{SON}{Self-Organizing Network}
\newacronym{sptcp}{SPTCP}{Single Path TCP}
\newacronym{srb}{SRB}{Service Radio Bearer}
\newacronym{srn}{SRN}{Standard Radio Node}
\newacronym{srs}{SRS}{Sounding Reference Signal}
\newacronym{ss}{SS}{Synchronization Signal}
\newacronym{sss}{SSS}{Secondary Synchronization Signal}
\newacronym{st}{ST}{Spanning Tree}
\newacronym{svc}{SVC}{Scalable Video Coding}
\newacronym{tb}{TB}{Transport Block}
\newacronym{tcp}{TCP}{Transmission Control Protocol}
\newacronym{tdd}{TDD}{Time Division Duplexing}
\newacronym{tdm}{TDM}{Time Division Multiplexing}
\newacronym{tdma}{TDMA}{Time Division Multiple Access}
\newacronym{tfl}{TfL}{Transport for London}
\newacronym{tfrc}{TFRC}{TCP-Friendly Rate Control}
\newacronym{tft}{TFT}{Traffic Flow Template}
\newacronym{tgen}{TGEN}{Traffic Generator}
\newacronym{tip}{TIP}{Telecom Infra Project}
\newacronym{tm}{TM}{Transparent Mode}
\newacronym{to}{TO}{Telco Operator}
\newacronym{tr}{TR}{Technical Report}
\newacronym{trp}{TRP}{Transmitter Receiver Pair}
\newacronym{ts}{TS}{Technical Specification}
\newacronym{tti}{TTI}{Transmission Time Interval}
\newacronym{ttt}{TTT}{Time-to-Trigger}
\newacronym{tue}{TUE}{Test UE}
\newacronym{tx}{TX}{Transmitter}
\newacronym{uas}{UAS}{Unmanned Aerial System}
\newacronym{uav}{UAV}{Unmanned Aerial Vehicle}
\newacronym{udm}{UDM}{Unified Data Management}
\newacronym{udp}{UDP}{User Datagram Protocol}
\newacronym{udr}{UDR}{Unified Data Repository}
\newacronym{ue}{UE}{User Equipment}
\newacronym{uhd}{UHD}{\gls{usrp} Hardware Driver}
\newacronym{ul}{UL}{Uplink}
\newacronym{um}{UM}{Unacknowledged Mode}
\newacronym{uml}{UML}{Unified Modeling Language}
\newacronym{upa}{UPA}{Uniform Planar Array}
\newacronym{upf}{UPF}{User Plane Function}
\newacronym{urllc}{URLLC}{Ultra Reliable and Low Latency Communications}
\newacronym{usa}{U.S.}{United States}
\newacronym{usim}{USIM}{Universal Subscriber Identity Module}
\newacronym{usrp}{USRP}{Universal Software Radio Peripheral}
\newacronym{utc}{UTC}{Urban Traffic Control}
\newacronym{vim}{VIM}{Virtualization Infrastructure Manager}
\newacronym{vm}{VM}{Virtual Machine}
\newacronym{vnf}{VNF}{Virtual Network Function}
\newacronym{volte}{VoLTE}{Voice over \gls{lte}}
\newacronym{voltha}{VOLTHA}{Virtual OLT HArdware Abstraction}
\newacronym{vr}{VR}{Virtual Reality}
\newacronym{vran}{vRAN}{Virtualized \gls{ran}}
\newacronym{vss}{VSS}{Video Streaming Server}
\newacronym{v2x}{V2X}{vehicle-to-everything}
\newacronym{v2i}{V2I}{vehicle-to-infrastructure}
\newacronym{v2v}{V2V}{vehicle-to-vehicle}
\newacronym{v2n}{V2N}{vehicle-to-network}
\newacronym{wbf}{WBF}{Wired Bias Function}
\newacronym{wf}{WF}{Waterfilling}
\newacronym{wg}{WG}{Working Group}
\newacronym{wlan}{WLAN}{Wireless Local Area Network}
\newacronym{osm}{OSM}{Open Source \gls{nfv} Management and Orchestration}
\newacronym{pnf}{PNF}{Physical Network Function}
\newacronym{drl}{DRL}{Deep Reinforcement Learning}
\newacronym{mtc}{MTC}{Machine-type Communications}
\newacronym{osc}{OSC}{O-RAN Software Community}
\newacronym{mns}{MnS}{Management Services}
\newacronym{ves}{VES}{\gls{vnf} Event Stream}
\newacronym{ei}{EI}{Enrichment Information}
\newacronym{fh}{FH}{Fronthaul}
\newacronym{fft}{FFT}{Fast Fourier Transform}
\newacronym{laa}{LAA}{Licensed-Assisted Access}
\newacronym{plfs}{PLFS}{Physical Layer Frequency Signals}
\newacronym{ptp}{PTP}{Precision Time Protocol}
\newacronym{lidar}{LiDAR}{Light Detection And Ranging}
\newacronym{dem}{DEM}{Digital Elevation Model}
\newacronym{dtm}{DEM}{Digital Terrain Model}
\newacronym{dsm}{DEM}{Digital Surface Models}
\newacronym{ota}{OTA}{Over-The-Air}
\newacronym{ns}{NS}{Network Slicing}
\newacronym{ne}{NE}{Nash Equilibrium}
\newacronym{hf}{HF}{High Frequency}
\newacronym{noma}{NOMA}{Non-Orthogonal Multiple Access}
\newacronym{sre}{SRE}{Smart Radio Environment}
\newacronym{ris}{RIS}{Reconfigurable Intelligent Surface}
\newacronym{inp}{InP}{Infrastructure Provider}
\newacronym{smf}{SMF}{Slicing Magangement Framework}
\newacronym{nsn}{NSN}{Network Slicing Negotiation}
\newacronym{sms}{SMS}{Slicing MAC Scheduler}
\newacronym{brd}{BRD}{Best Response Dynamics}
\newacronym{dssbr}{DSSBR}{Double Step Smoothed Best Response}
\newacronym{poa}{PoA}{Price of Anarchy}
\newacronym{pos}{PoS}{Price of Stability}
\newacronym{milp}{MILP}{Mixed Integer-Linear Program}
\newacronym{pod}{PoD}{Price of DSSBR}
\newacronym{roc}{ROC}{Radio Overload Control}
\newacronym{ciot}{cIoT}{critical Internet of Things}
\newacronym{embbpr}{eMBB Pr.}{enhanced Mobile BroadBand Premium}
\newacronym{sps}{SPS}{Semi-persistent Scheduling}
\newacronym{cg}{CG}{Configured Grant}
\newacronym{embbbs}{eMBB Bs.}{enhanced Mobile BroadBand Basic}
\newacronym{en}{EN}{Edge Node}
\newacronym{ec}{EC}{Edge Computing}
\newacronym{sp}{SP}{Service Provider}
\newacronym{me}{ME}{Market Equilibrium}
\newacronym{so}{SO}{Social Optimum}
\newacronym{wso}{WSO}{Weighted Social Optimum}
\newacronym{ps}{PS}{Proportional Sharing}
\newacronym{eg}{EG}{Eisenberg-Gale program}
\newacronym{pe}{PE}{Pareto Efficiency}
\newacronym{nsw}{NSW}{Nash Social Welfare}
\newacronym{ef}{EF}{Envy-Freeness}
\newacronym{sub6}{sub6GHz}{Below 6GHz}
\newacronym{ncr}{NCR}{Network-Controlled Repeater}
\newacronym{nlos}{NLoS}{Non-Line of Sight}
\newacronym{src}{SRC}{Smart Radio Connection}
\newacronym{srd}{SRD}{Smart Radio Device}
\newacronym{cs}{CS}{Candidate Site}
\newacronym{tp}{TP}{Test Point}
\newacronym{fov}{FoV}{Field of View}
\newacronym{nrric}{near-RT RIC}{Near Real-time {RAN} Intelligent Controller}
\newacronym{e2ap}{E2AP}{E2 Application Protocol}
\newacronym{e2sm}{E2SM}{E2 Service Model}
\newacronym{nrtric}{non-RT RIC}{Non-Real-Time {RIC}}
\newacronym{itti}{ITTI}{Inter-task Interface}
\newacronym{bap}{BAP}{Backhaul Adaptation Protocol}
\newacronym{iabest}{IABEST}{Integrated Access and Backhaul Experimental large-Scale Tetbed}
\newacronym{teid}{TEID}{Tunnel Endpoint Identifier}
\newacronym{dlsch}{DL-SCH}{Downlink Shared Channel }
\newacronym{ulsch}{UL-SCH}{Uplink Shared Channel }
\newacronym{rsu}{RSU}{Road Side Unit}
\newacronym{its}{ITS}{Intelligent Transportation Systems}
\newacronym{vanet}{VANET}{Vehicular Ad-hoc Network}
\newacronym{dt}{DT}{Digital Twin}
\newacronym{ecc}{ECC}{Edge Computing Cluster}
\newacronym{o2i}{O2I}{Outdoor-to-indoor}
\newacronym{fwa}{FWA}{Fixed wireless access}
\newacronym{afc}{AFC}{Automated Frequency Coordinator}
\newacronym{bb}{BB}{baseband}
\newacronym{cpri}{CPRI}{Common Public Radio Interface}
\newacronym{ecpri}{eCPRI}{Enhanced CPRI}
\newacronym{re}{RE}{Resource Element}
\title{Shaping Radio Access to Match Variable Wireless Fronthaul Quality in Next-Generation Networks}  %

\author{\IEEEauthorblockN{Marcello Morini\textsuperscript{1}, Eugenio Moro\textsuperscript{1}, Ilario Filippini\textsuperscript{1}, Danilo De Donno\textsuperscript{2}, Antonio Capone\textsuperscript{1}}
\IEEEauthorblockA{\textsuperscript{1} DEIB, Politecnico di Milano, Milan, Italy - \textit{\{name.surname\}@polimi.it}}
Milan Research Center, Huawei Technologies Italia S.r.l., Milan, Italy - \textit{danilo.dedonno@huawei.com}\vspace{-1.5em}}

\maketitle

\begin{abstract}

The emergence of \gls{cran} has revolutionized mobile network infrastructure, offering streamlined cell-site engineering and enhanced network management capabilities. As \gls{cran} gains momentum, the focus shifts to optimizing fronthaul links. While fiber fronthaul guarantees performance, wireless alternatives provide cost efficiency and scalability, making them preferable in densely urbanized areas. However, wireless fronthaul often requires expensive over-dimensioning to overcome the challenging atmospheric attenuation typical of high frequencies. We propose a framework designed to continuously align radio access capacity with fronthaul link quality to overcome this rigidity. By gradually adapting radio access capacity to available fronthaul capacity, the framework ensures smooth degradation rather than complete service loss. Various strategies are proposed, considering factors like functional split and beamforming technology and exploring the tradeoff between adaptation strategy complexity and end-to-end system performance.
Numerical evaluations using experimental rain attenuation data illustrate the framework's effectiveness in optimizing radio access capacity under realistically variable fronthaul link quality, ultimately proving the importance of adaptive capacity management in maximizing C-RAN efficiency.
\end{abstract}

\IEEEpeerreviewmaketitle

\vspace{-0.5em}
\section{Introduction}
\glsresetall
\gls{cran} represents the culmination of the device disaggregation process in \glspl{ran}. This architecture involves concentrating the \gls{bb} processing of a cluster of cells in a central location and connecting them to \glspl{ru} dislocated at the cell sites through high-capacity, low-latency connections, called fronthaul links. This approach has gained significant momentum with the introduction of 5G technologies, which made virtualization techniques more accessible. 
Recent global market surveys conducted among mobile radio network operators indicate that 80\% of the respondents plan to implement \gls{cran} in at least 20\% of their sites (with nearly half of them targeting more than 40\% of sites) by the end of 2025, a notable increase from 2022, when the operators aiming for this level of site coverage were 40\% \cite{lightreading_cran}.  

The appeal of \gls{cran} architecture lies in its ability to streamline cell-site engineering and reduce the geographical distribution of maintenance sites. Centralizing \gls{bb} processing simplifies the design of \glspl{ru} in a cost-effective and energy-efficient manner. Moreover, it enhances network coordination and management, enabling better strategies for load balancing, coordinated multi-point communications, cooperative spatial multiplexing, macro-diversity, and mobility management. Additionally, \gls{cran} facilitates network virtualization, slicing, and openness \cite{7064897}.

The primary technical challenge in \gls{cran} deployment is designing fronthaul links with sufficient capacity and minimal latency. Two options are available: fiber fronthaul and wireless fronthaul. Fiber fronthaul guarantees high capacity and minimal latency but can be cost-prohibitive due to high trenching expenses, particularly in urban European scenarios~\cite{lashgari2022fiber}. Alternatively, wireless fronthaul provides cost efficiency and allows for rapid, flexible, and scalable deployment, especially in areas where fiber connectivity is impractical or unavailable.

Realizing the full potential of wireless fronthaul relies on establishing high-capacity and reliable links. Commonly utilized technologies for achieving this operate within frequency bands such as D, K, E, V, and W. However, communications at these frequencies are prone to significant propagation losses and are vulnerable to various obstacles and weather conditions. For instance, rainfall can impact the link's power budget, introducing substantial additional attenuation.

Fronthaul and radio access capacities are strongly interdependent. Traditional fronthaul links are engineered to maintain a constant capacity, capable of providing to the \gls{ru} the highest amount of resources (e.g., \gls{rb},  \acrshort{mimo} layers) it needs to achieve the maximum data rate in access. However, if fronthaul capacity drops below a certain threshold, the configured access scheme becomes unsustainable and the entire cell drops. Overdimensioning the link budget may be a reasonable solution for wired fronthaul, but it is not efficient nor sustainable for its wireless counterpart. 

In response to variable wireless fronthaul link quality, a more effective strategy involves adaptively reducing the required fronthaul rate by gradually degrading \gls{e2e} radio access performance. Although temporarily reducing the data rate is preferable to losing a cell entirely, quantifying the impact of access capacity degradation on fronthaul capacity is not straightforward. It hinges on factors like the employed functional split, the cell configuration and the type of beamformer used. Moreover, different radio access throttling countermeasures can be used to mitigate fronthaul link requirements, each with its own advantages and drawbacks depending on the operational scenario. Selecting the optimal scheme is critical to adapt to the available capacity effectively.

This paper introduces a modeling framework designed to align radio access capacity with fronthaul link quality. This framework enables us to estimate the upper bound of radio access capacity achievable with any given fronthaul capacity. Through this, we explore all potential countermeasures to reduce fronthaul link requirements when the highest capacity is unavailable, highlighting their benefits and limitations. Finally, we present numerical examples of variable-quality fronthaul links and their achievable radio access capacity using experimental attenuation time series collected during a rainfall event. These examples demonstrate how radio access capacity throttling can be performed based on available fronthaul capacity.

The rest of the paper is structured as follows. In Sec.~\ref{sec:background}, we discuss \gls{cran} architectures and components, while in Sec.~\ref{sec:model} we introduce our framework to model fronthaul and radio access capacity. Sec.~\ref{sec:countermeasures} discusses all potential countermeasures for throttling radio access capacity to adapt to the available fronthaul link quality, and numerical examples are given in Sec.~\ref{sec:numerical}. Finally, Sec.~\ref{sec:conclusions} concludes the paper.

\vspace{-0.5em}
\section{Background} \label{sec:background}
In \gls{cran}, the fronthaul plays a crucial role as the connection layer between a \acrfull{bb} central unit and a collection of \glspl{ru}. This design involves the geographic separation and the disaggregation of traditional RAN cell sites into distinct network functions, adhering to a paradigm where \glspl{ru}, at the far edge of the \gls{ran}, primarily handle radio signal reception and transmission at the cell site. Meanwhile, all other functions are centralized within the unit at the opposite endpoint – the proximal endpoint – of the backhaul link. %

\begin{figure}
    \centering
    \includegraphics[width=0.8\columnwidth]{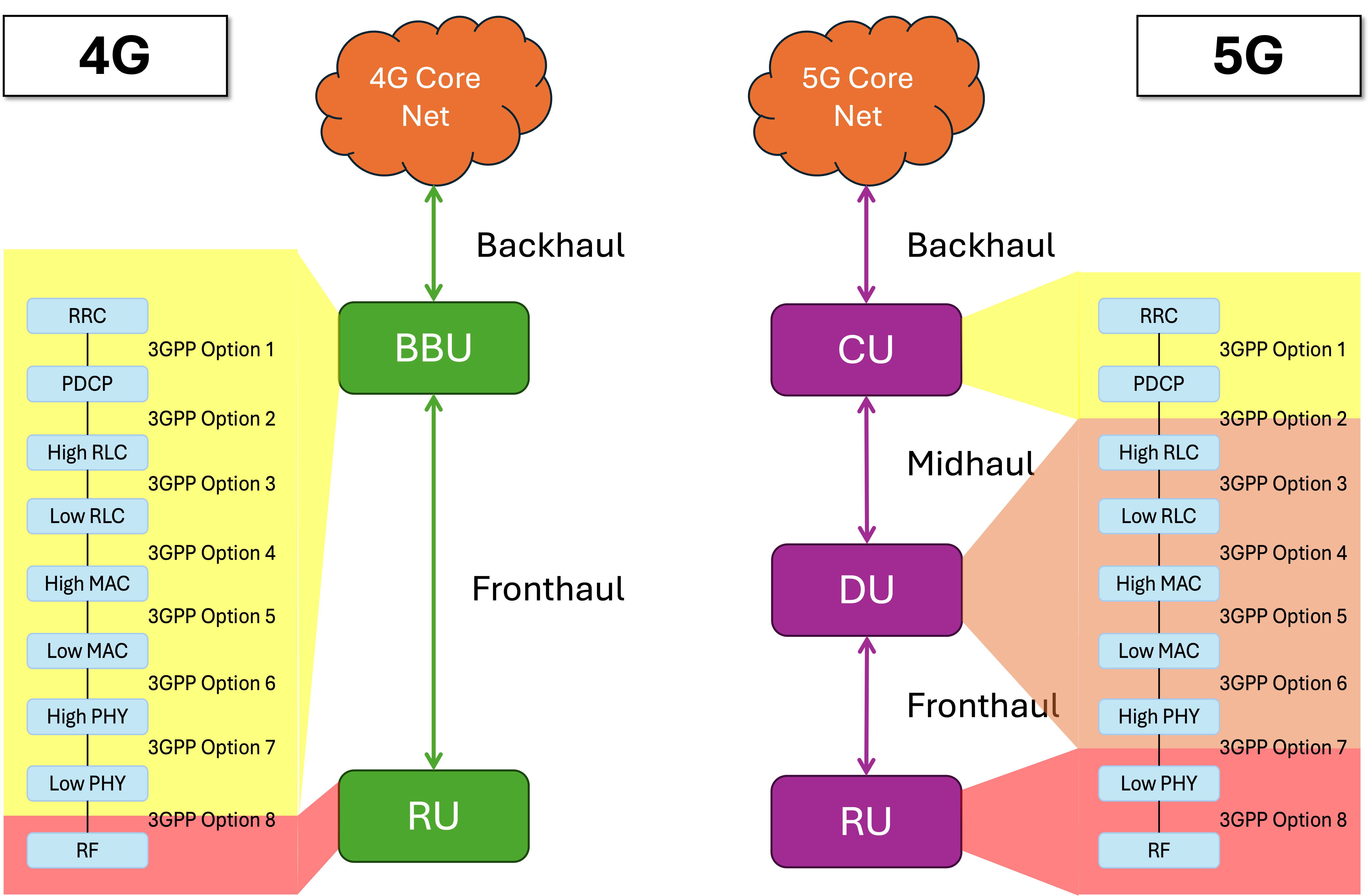}
    \caption{\small 3GPP 4G vs 5G RAN architectures and functional split options
    \vspace{-1.5em}
    }
    \label{fig:fronthaul_arch}
\end{figure}

As depicted in Figure~\ref{fig:fronthaul_arch}, the role of the proximal endpoint has evolved with each generation of mobile communications technology. In 4G \glspl{ran}, the architecture comprises \glspl{bbu} and \glspl{ru} interconnected via the fronthaul. In 5G specifications, an additional functional split has been introduced, where real-time signaling procedures are managed by \glspl{du} connected via midhaul links, while non-real-time higher-layer protocol functions are overseen by the \gls{cu}.

A more precise delineation of the functions separated by a fronthaul link is provided by 3GPP \gls{ran} functional split options, also illustrated in Figure~\ref{fig:fronthaul_arch}. The rationale underlying these options is that the lower the split level, the lesser the signal/data processing is performed at the \gls{ru}. Consequently, more raw data is transmitted through fronthaul links, resulting in higher throughput.

In today's landscape, Option 7 for fronthaul links stands out as one of the most favored choices for 5G networks, also supported by the \gls{oran} Alliance. This option entails hosting high and low physical (PHY) layer functions at the two endpoints of the fronthaul links. PHY-layer functions are applied to convert transport blocks received from the MAC layer into IQ samples ready for the RF frontend. Within the PHY layer, a more refined functional split can be delineated, allowing for an asymmetric split between uplink and downlink function chains. Two common examples are Option 7-3, which splits the PHY layer between the \textit{scrambling} and the \textit{modulation} functions, thus transporting codewords over the fronthaul link, or Option 7-2, which makes a split between the \textit{precoding} and the \textit{resource element mapping} functions, therefore sending antenna symbols through the fronthaul. 

The protocols utilized for the fronthaul interface have evolved over the past decade. Initially, in the early 4G systems, the \gls{bbu}-\gls{ru} interface was proprietary to mobile equipment vendors and built upon the \gls{cpri} interface, implementing split Option 8. However, in 2017, an updated interface known as \gls{ecpri} was introduced. The \gls{ecpri} interface offers remarkable flexibility and supports multiple split options, all aligned with the 3GPP RAN functional split \cite{eCPRI_standard}. Consequently, it quickly emerged as the standard interface linking \gls{ru} and \gls{du} in 5G networks.

Wireless fronthaul capacity issues have been known since \gls{cran} architectures were first discussed several years ago. Still, we believe that our contributions stand out in several ways from the limited, existing literature. Preliminary fronthaul rate formulas were reported in \cite{fh_dimensioning_2}. That contribution, however, focuses on wired connections and accounts for only a subset of splits. A survey of possible splits and their required rate and latency is provided in \cite{fh_dimensioning_1}. However, the proposed formulae and data are outdated and necessitate a revision. The recent survey in \cite{challenges_opportunities_wfh} discusses the challenges to be faced when operating a wireless fronthaul, touching on data rate, latency, frame loss, and jitter. Nonetheless, it does not touch on the critical role of the beamforming control information, which may significantly impact the required fronthaul capacity. Also, it lacks a general model for the fronthaul rate calculation.
Several ways to answer to wireless fronthaul challenges are mentioned in \cite{wfh_for_5g, fh_constrained_cran_survey, flexible_cran}, but without proposing 
specific methodologies or numerically validating any implementation.
Finally, let us point out that our contribution substantially differs from the set of network planning works related to \textit{joint access and fronthaul optimization}, as we propose online reconfiguration strategies that can be easily implemented in greenfield scenarios and already deployed networks.

\vspace{-0.25em}
\section{Analytical model}  \label{sec:model}
This section presents a mathematical model of the fronthaul rate for arbitrary cell configuration parameters and different split options of the PHY layer. 

Generally speaking, the overall traffic transported by the fronthaul link can be split into 2 components: data directed to the radio access interface to be transmitted by the \gls{ru} (i.e., unmodulated bits or IQ samples), and antenna beamformer control data directed to the \gls{ru}'s \gls{rf} chains. The additional data volume required by the synchronization and management planes is not considered in our analysis, as the corresponding volumes are negligible with respect to the overall fronthaul capacity.

\begin{figure}  
    \centering
    \includegraphics[width=0.80\linewidth]{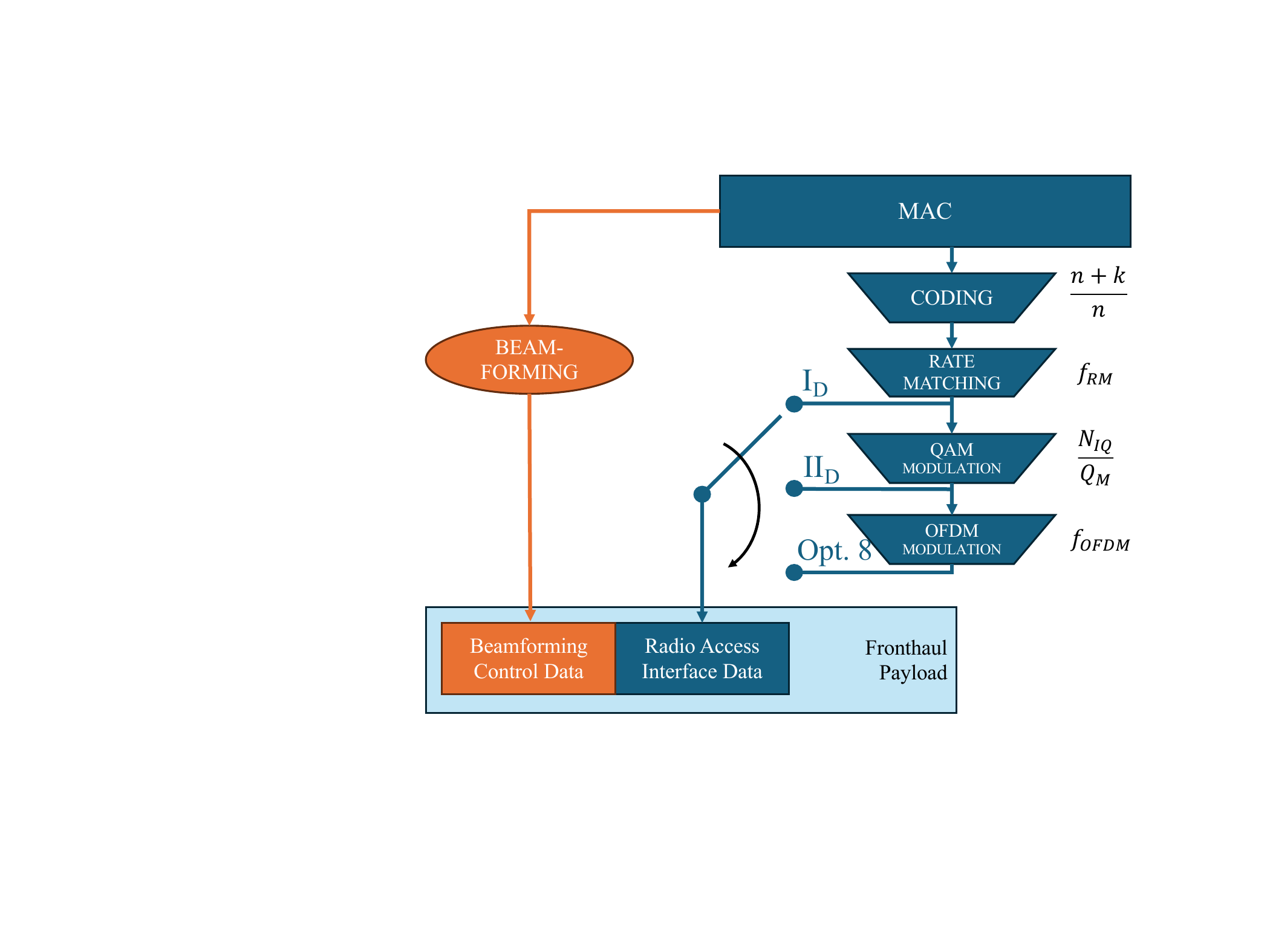}
    \caption{\small Downlink physical layer functional blocks impacting the payload \vspace{-1.5em}}
    \label{fig:usefulPhyBlocks}
\end{figure}

The beamforming control rate depends only on the beamforming technology used by the \gls{ru} and it is independent on the selected split, as shown in Fig.~\ref{fig:usefulPhyBlocks}.

On the other hand, the radio access interface data is strongly related to the selected split, as this corresponds to the PHY payload size transported on the fronthaul link. 
Consider Fig.~\ref{fig:usefulPhyBlocks} again. On the right side are reported only the downlink PHY functional blocks that modify the size of the payload transported over the fronthaul link and \textit{some} of the most used splits, suggested by ~\cite{eCPRI_standard}. 
Next to each block of the radio access interface chain, the impact of the performed function on the payload is indicated by means of specific multiplicative factors. Those factors must be sequentially multiplied to the payload dimension when blocks are crossed from the top to the bottom, and divided otherwise. 
For instance, if \textit{split $I_D$} is selected, the dimension of the payload coming from the MAC has to be multiplied by the coding factor $\frac{n+k}{n}$, where $n$ is the original size and $k$ are the additional redundancy bits, 
and the rate matching factor $f_{RM}$.

While this method can be applied regardless of the direction of access transmission, our analysis focuses on downlink for the sake of simplicity.
\subsection{Radio access interface rate}
We now detail the mathematical model that allows us to compute the fronthaul volume contribution given by the radio access interface data.
Starting from the bottom part of the right branch in Fig.~\ref{fig:usefulPhyBlocks}, we consider the time-frequency-space resource grid situated in correspondence of eCPRI \textit{split $II_D$} (equivalent to 3GPP Option 7-2). 
Each \gls{re} of each MIMO layer can carry one IQ sample. Considering each IQ sample is encoded by using $N_{IQ}$ bits, and that the symbol transmission time is equal to one OFDM symbol duration $T_S$, we obtain the \textit{Split $II_D$} fronthaul rate formula:
\begin{equation}
R_{FH}^{II_D} =  N_{RB} \cdot N_{SC} \cdot  N_{MIMO} \cdot N_{IQ} \cdot\frac{1}{T_S}.    \label{R_II_D}
\end{equation}
Here $N_{RB}$ is the number of \glspl{rb} available for a given channel bandwidth and subcarrier spacing, as defined in \cite{TS38.104}. Multiplying this with the number of subcarriers per \gls{rb}, namely $N_{SC}$, we obtain the total number of \gls{re}. $N_{MIMO}$ is the number of downlink TX antenna ports, corresponding to the maximum number of concurrently active MIMO layers. Note that, although not represented in Fig.~\ref{fig:usefulPhyBlocks}, this rate equals the fronthaul rate of the uplink \textit{Split $I_U$}.

To obtain the fronthaul rate of \textit{Split $I_D$} (equivalent to 3GPP Option 7-3), it suffices to move up by one functional block in Fig.~\ref{fig:usefulPhyBlocks} and compute the rate between \textit{QAM modulation} and \textit{Rate matching} blocks. The order-$M$ modulation operation transforms sets of $Q_M=2^M$ bits in sets of IQ samples, represented with $N_{IQ}$ bits; thus the factor of this block is $\frac{N_{IQ}}{Q_M}$. Dividing the Eq.~\ref{R_II_D} by such factor, we obtain the following:
\begin{equation}
R_{FH}^{I_D} = N_{RB} \cdot N_{SC} \cdot N_{MIMO} \cdot Q_M \cdot\frac{1}{T_S} \label{R_I_D}
\end{equation}

Considering splits above the \textit{Rate matching} block is possible but challenging as CRC calculation, channel coding, and rate matching differ for each single physical channel (e.g., PBCH, PDSCH, PDCCH...).
Split options below \textit{Split $II_D$} require to include the OFDM factor $f_{OFDM}$. The OFDM factor encloses two contributions: one for the frequency-domain zero padding performed over unused \glspl{re}, and one for the cyclic prefix insertion performed in the time domain.

The total radio access rate the fronthaul link must support depends on the specific \gls{fdd} or \gls{tdd} configuration.
If the system is \gls{tdd}, the fronthaul must support the largest access rate between the uplink and the downlink.
For the \gls{fdd} case, the uplink and the downlink channels will have separate \gls{rb} allocations (resulting in different values of $N_{RB}$) and potentially different MIMO schemes. 
In this case, the minimum radio access interface rate corresponds to the sum of the uplink and downlink rates computed according to the equations above.

\subsection{Beamforming control rate}
The left branch in Fig. \ref{fig:usefulPhyBlocks} accounts for the main antenna control data generated at the DU-side to be transported over the fronthaul: the beamforming weights. Generally speaking, the beamforming control data varies significantly depending on the beamforming technique in use.

In the case of \gls{abf}, the fronthaul link transports $N_{ANT}$ phase-shift coefficient, that is, one per each antenna element. Consequently, the \gls{abf} control rate can be computed as follows:
\begin{equation}
R_{FH}^{ABF} = N_{ANT} \cdot b_{PS} \cdot \frac{1}{T_S}, \label{R_ABF}
\end{equation}
where $b_{PS}$ bits are used to encode the coefficients to be applied at the $N_{ANT}$ antennas every $T_S$ seconds, namely the symbol time\footnote{The choice of $T_S$ is technologically challenging for the beamforming hardware, but conservative with respect to the final required fronthaul capacity.}.
As it will be numerically shown in Sec.~\ref{sec:numerical}, the resulting rate is significantly lower than the radio access interface rate.

Inversely, \gls{dbf} control rate is typically significantly larger. This technology guarantees more accurate antenna radiation patterns by digitally applying amplitude and phase corrections to a large set of signals generated by multiple RF chains. Moreover, different beamforming can be performed across the entire band by applying different weights to every frequency element. However, the increased overall system performance and flexibility come at a higher transport cost. Indeed, the total \gls{dbf} data rate can be computed as
\begin{equation}
R_{FH}^{DBF} = N_{FE} \cdot N_{MIMO} \cdot N_{RFC} \cdot b_W \cdot \frac{1}{T_S}
\label{R_DBF_FH}
\end{equation}
where $N_{FE}$ is the number of frequency elements that can be controlled (i.e., the control granularity over the frequency resource), $N_{RFC}$ is the number of \gls{rf} chains, $b_W$ is the bit width of the precoding weights (i.e., the beamforming weights). Here, a reasonable value for $N_{FE}$ may be $N_{RB}$, while a beamforming switch time of $T_S$ is considered, similarly to the \gls{abf} case.\\ %
Differently from \gls{abf}, the final control data rate can approach the radio access interface rate, thus having a significant impact on the required fronthaul capacity.

Finally, the control rate for Hybrid Beamforming by summing Eqs.~\ref{R_ABF} and~\ref{R_DBF_FH} with the proper number of \gls{rf} chains and antennas.

\subsection{Cell capacity}
The formula to compute the radio access capacity of a single RU is provided by \gls{3gpp} \cite{TS38.306}. We report it here with the notation employed in this paper:
\begin{equation}
R_{ACC} = N_{RB} \cdot N_{SC} \cdot N_{MIMO} \cdot Q_M \cdot R_{MAX} \cdot \frac{(1 - OH)}{T_S}.  \label{R_ACC}   
\end{equation}

It derives directly from Eq. \eqref{R_I_D} and includes the value of the target code rate, $R_{MAX}$, and the overhead due to the control channels, $OH$, (e.g., PBCH, PRACH). Consequently, it indicates the maximum cumulative effective data rate achievable by the users connected to the antenna. One more factor $f_{TDD}$ should be multiplied in case of \gls{tdd} adoption, to explicit the portion of resources devoted to uplink and to downlink.

\section{Capacity reduction countermeasures}    \label{sec:countermeasures}
Unforeseen channel conditions can reduce the fronthaul link capacity below the threshold required to support the cell. In this case, the communication between the \gls{du} and the \gls{ru} cannot be guaranteed, and the system is considered to be in an outage. However, there exists possible countermeasures to adapt the fronthaul link rate to the instantaneous link capacity variations.

As shown in Sec.~\ref{sec:model}, the required fronthaul rate is a function of the cell configuration and allocated resources. As such, we can act on these to design countermeasures to fronthaul capacity degradation. In this section, we propose two strategies, for which we analyze their advantages and drawbacks and discuss their expected performance. These two strategies will be then applied to real-world measurements of a fronthaul link in Sec.~\ref{sec:numerical} to numerically evaluate the performance in terms of \gls{e2e} cell capacity.

\subsection{\gls{cr}}
According to Eq.~\eqref{R_II_D}, the fronthaul capacity required to support the radio access interface data rate is directly proportional to the configured cell resources. In particular, this rate increases linearly with the bandwidth and the number of MIMO layers. As such, a straightforward adaptation strategy could consist of an on-the-fly reconfiguration of these cell parameters. For instance, an \gls{fr2} cell working with a bandwidth of $200$ MHz could be temporarily reconfigured to use only $100$ MHz, yielding a fronthaul rate reduction of a factor 2. A similar effect can be achieved by deactivating MIMO layers. 

Concerning the beamforming control data, \textit{cell reconfiguration} does not affect the capacity requirements when \gls{abf} is used. On the other hand, the data rate for \gls{dbf} is once again directly proportional to the bandwidth and number of MIMO layers, as shown in Eq.~\eqref{R_DBF_FH}. As such, in the case of hybrid or digital beamforming, the \textit{cell reconfiguration} strategy can effectively reduce the fronthaul traffic by acting on the same parameters.
\subsection{\gls{sbt}}

\textit{Cell reconfiguration} is a straightforward way of controlling the fronthaul traffic. However, as discussed later in the text, this strategy shows some crucial limitations. For this reason, we propose \textit{scheduler-based throttling} as an alternative adaptation method. 

Its rationale is shared with cell reconfiguration, but the factors defining the required fronthaul rate in Eqs.~\eqref{R_II_D} and~\eqref{R_DBF_FH} are controlled at the radio resource management level. Indeed, the resource scheduling process can be instructed to avoid allocating a certain portion of radio resources to reduce the fronthaul capacity requirements. 

For instance, if a contiguous portion of \glspl{rb} is marked as never to be allocated, these do not need to be transported over the fronthaul link. Alternatively, if a subset of MIMO layers is not used to schedule transmissions, the relative fronthaul capacity can be spared. In both cases, the effect is equivalent to a cell reconfiguration, although with increased granularity with respect to the relatively limited set of legal cell parameters. 

\subsection{Strategy comparison}
While the two strategies detailed above can effectively adapt the fronthaul traffic to the available capacity, they show some differences in complexity and performance. 

\textit{Cell reconfiguration} shows the lowest implementation and execution complexity. Indeed, cell reconfiguration is a natural operation for base stations. \textit{Scheduler-based throttling}, on the other hand, requires a custom scheduler that can be controlled based on the instantaneous fronthaul capacity. \textit{Cell reconfiguration} also appears to be compatible with any split, while scheduling control might not be able to throttle the fronthaul traffic for very low splits (e.g., \textit{Split E}) where the baseband signal is exchanged on the fronthaul link. 

Nevertheless, the increased complexity of the \textit{scheduler-based throttling} comes with higher adaptation performance. By controlling the scheduling decision, the fronthaul traffic can be throttled with a granularity equivalent to the capacity required to transport a single \gls{rb}. Consequently, the gap between the fronthaul rate and the available capacity can be reduced to a minimum, resulting in higher access performance. 

Cell parameter reconfiguration is generally not designed to be an instantaneous operation that can be carried out frequently. Therefore, the reaction speed of the \textit{cell reconfiguration} approach is limited. Additionally, any reconfiguration might disrupt end-user connectivity, further increasing the cost of applying the strategy. Conversely, scheduling policies can be changed to a \gls{tti} notice, and they do not cause connectivity disruptions other than the necessary radio access capacity reduction.  

\section{Numerical Analysis} \label{sec:numerical}
\begin{figure}
    \vspace{0.04in}
    \centering
    \begin{subfigure}{1\linewidth}
        \centering
        \includegraphics[width=0.9\linewidth, trim={0 0.5cm 0 0},clip]{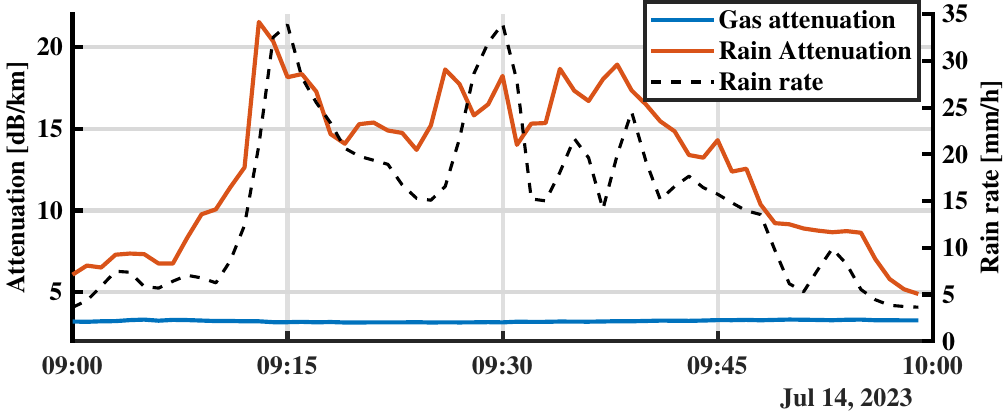}
        \caption{Attenuation and rain rate}
        \label{fig:attenuationVsTime}
    \end{subfigure}

    \begin{subfigure}{\linewidth}
        \centering
        \includegraphics[width=0.9\linewidth, trim={0 0.5cm 0 0},clip]{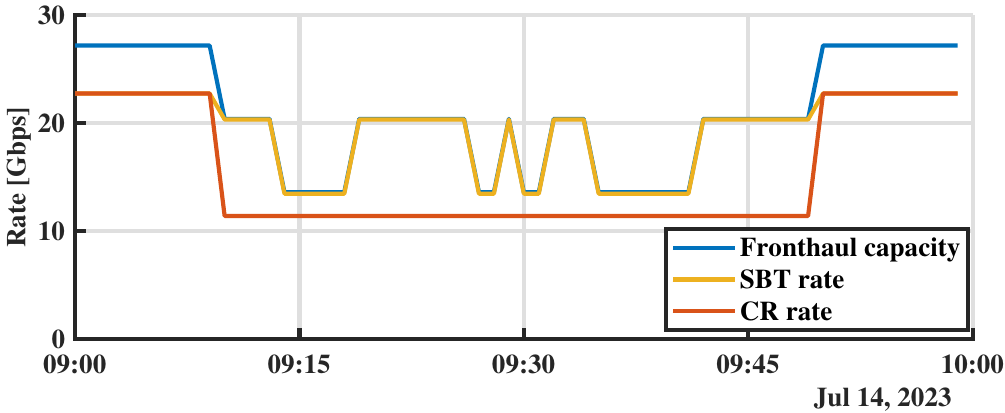}
        \caption{Channel capacity and adapted rates}
        \label{fig:capacityInTime}
    \end{subfigure}

    \begin{subfigure}{\linewidth}
        \centering
        \includegraphics[width=0.9\linewidth, trim={0 0.5cm 0 0},clip]{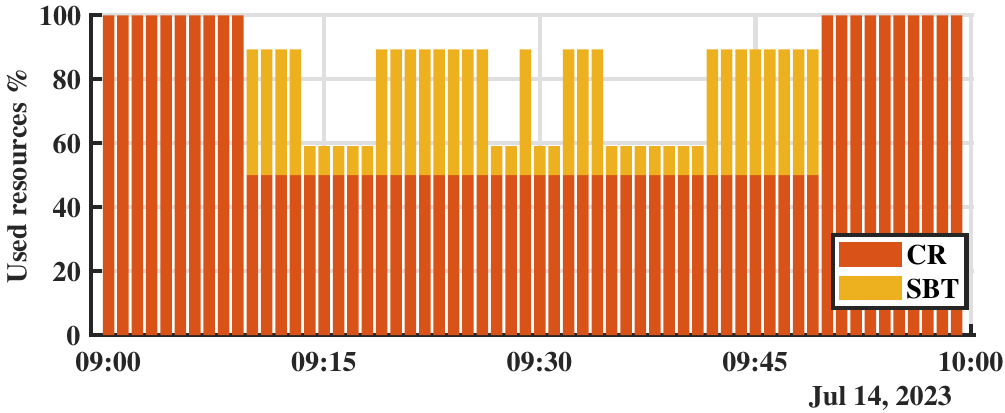}
        \caption{\% of used resources ($N_{RB}\cdot N_{MIMO}$)}
        \label{fig:RBinTime}
    \end{subfigure} %

    \caption{Time-domain plots of the analyzed scenario}
\end{figure}
\begin{table}[]
    \vspace{0.04in}
    \caption{D-band fronthaul parameters}
    \begin{center}
    \begin{tabular}{|c|c|}
    \hline
    \textbf{Distance} & 1 km    \\ \hline
    \textbf{EIRP}     & 65 dBm  \\ \hline
    \textbf{RX antenna gain}     & 42 dBi  \\ \hline
    \textbf{Free-space path loss}     & 137 dB  \\ \hline
    \textbf{Available \gls{fh} modulations}     & QPSK, 64QAM, 256QAM  \\ \hline
    \textbf{Available FH BW}     & 250, 500, 1000, 2000 MHz  \\ \hline
    \end{tabular}
    \label{tab:D_band_link_param}
    \end{center}
    \vspace{-0.2in}
\end{table}

In this section, we evaluate the impact of the strategies outlined in Sec. \ref{sec:countermeasures} by applying them to real, measured data as those provided in  \cite{attenuation_verdugo}. We begin by discussing measurements and capacity aspects of an experimental wireless link used as a fronthaul candidate. Subsequently, we analyze the results of applying the reconfiguration strategies to this link.

Authors in \cite{attenuation_verdugo} measure the attenuation over time along a wireless link observed every second for a whole day in the summer of 2023, focusing on both rain and gas (e.g., fog and atmosphere) attenuation. The time series was measured over an E-band link (83 GHz) and then fed into a frequency scaling model to upconvert attenuation measurements to the D-band (156 GHz), more suitable for fronthaul transmissions. A subset of the results are shown in Fig. \ref{fig:attenuationVsTime}. 
The gas attenuation remains relatively constant over the day, at around $3 dB/km$. 
The rain attenuation is highly influenced by an intense rain event that happened from 9 to 10 AM, the portion of day shown in the figure. The precipitation rapidly increases up to $34 mm/h$, therefore classified as a \textit{very heavy rain} event. Such events are not common at the location of the experiment. Indeed, they occur only $0.01\%$ of the time, a small probability that however cannot be neglected to guarantee a five-nines uptime availability.

We use these rain and gas attenuation values to run a model for computing the capacity of a realistic link working at the D-band undergoing the same rain event. The operating parameters of this link are reported in Tab. \ref{tab:D_band_link_param}. The link transmitter adapts its modulation depending on predefined SINR thresholds detected at a beacon receiver, eventually translating the overall attenuation in up to twelve fronthaul link capacity values. Among them, only three are triggered in the considered dataset: $27.2$  Gbps, $20.4$ Gbps and $13.6$ Gbps. The last two values are experimented only during the measured precipitation event, while the first is seen during the rest of the day. In this work, we assumed that only atmospheric attenuation influences the fronthaul link capacity. However, the discussion can be generalized by considering all other variable parameters in the link budget.

In the next step, we fed the time-varying fronthaul link capacity during the rain event into a simulator implementing both the analytical framework described in Sec. \ref{sec:model} and the adaptation strategies discussed in Sec. \ref{sec:countermeasures}. 
The goal of the simulator is to fill the available fronthaul link \textit{capacity} with the largest fronthaul \textit{rate} possible to provide the largest radio access interface capacity during the rain event. Without loss of generality, the modeled radio access interface considers a \gls{mmwave} RU, implementing up to 200 MHz of bandwidth (i.e., 132 RB), 8 \gls{mimo} layers, managed in a \gls{tdd} fashion, controlled by \acrfull{abf}, and using \textit{Split $II_D$}. Therefore, the total fronthaul rate is computed as the sum of Eqs. \eqref{R_II_D} and \eqref{R_ABF}. In the system we considered, we further assume that $N_{IQ} = 16 bits$, $N_{ANT} = 1024$ and $b_{PS} = 5 bits$, realistic assumptions for a \gls{mmwave} cell. Fronthaul and \gls{abf} rates reach up to 22.7 Gbps and 573 Mbps, respectively.

The link capacity over time and the rate resulting from the countermeasures application are reported in Fig. \ref{fig:capacityInTime}. We notice that the fronthaul capacity switches to lower modulation orders just after the precipitation starts.%

We test the two types of countermeasures. 
When the \textit{cell reconfiguration} strategy is used, we obtain the behavior reported in orange in Fig. \ref{fig:capacityInTime}. The system, starting from 200MHz and 8 \gls{mimo} layers, halves the bandwidth as the fronthaul capacity is reduced, and keeps this configuration until the end of the rain event. An equivalent effect could be obtained by halving the maximum number of \gls{mimo} layers. 
In case the scheduler can be accessed and programmed (i.e., \textit{scheduler-based throttling}), we get the behavior described in Fig. \ref{fig:capacityInTime} by the yellow curve, which is almost indistinguishable from the blue one of the link capacity. Varying the radio access capacity at steps of a single \gls{rb} (which corresponds to 0.76\% of the overall capacity), this method has much finer granularity in tuning radio access capacity. Indeed, the yellow curve closely approaches the available fronthaul capacity curve from below. 

The improved flexibility of \textit{scheduler-based throttling} is also highlighted with the cumulative bar plot in Fig.~\ref{fig:RBinTime}, where the percentage of resources used (intended as $N_{RB}\cdot N_{MIMO}$) is reported. As shown, the \textit{Scheduler based throttling} allows us to use up to 87\% of the total available resources, instead of the 50\% reached by the \textit{Cell reconfiguration}. 

The maximum radio access capacity can be computed  from Eq. \eqref{R_ACC}, with $R_{MAX} = 948/1024$, $OH = 0.18$ for \gls{dl} and $f_{TDD}^{DL} = 0.8$. Since it is directly proportional to the fronthaul rate, their trends are equal. 
Cell downlink access capacity varies from a maximum of 6.9 Gbps to a minimum of 3.4 Gbps in case of Cell reconfiguration, while Scheduler-based Throttling is capable of guaranteeing 6.9 Gbps, 6.0 Gbps and 3.9 Gbps.

Finally, we further illustrate the benefits of applying fronthaul adaptation strategies by comparing it to a more traditional system where no adaptation happens. In the latter case, a conservative choice implies over-provisioning the fronthaul link to guarantee an availability equal to 100\%, but at the cost of operating with the minimum reachable rate (i.e., $11.9$ Gbps, half of the maximum value). Otherwise, one can set the system to work at the maximum rate but accept 40 minutes of downtime (i.e., $97.2\%$ uptime availability). Both approaches appear to be relatively inefficient when compared to the adaptation techniques.
In particular, when \textit{scheduler-based throttling} is applied, it is possible to extract the best rate value in every condition, obtaining $23.3$ Gbps for $97.2\%$ of the day, $20.4$ Gbps for $1.6\%$ of the day and the minimum capacity of $13.5$ Gbps only for the remaining $1.1\%$ of the day. 
This approach turns the availability value into a step function, meaning that the cell rate is reconfigured according to the available channel capacity such that cumulative availability reaches 100\%.

\section{Conclusions} \label{sec:conclusions}
\gls{cran} is emerging as a popular architectural solution among mobile radio network operators. Leveraging a wireless fronthaul, \gls{cran} offers additional flexibility, cost-efficiency, and scalability advantages on top of the benefits of a fiber fronthaul. However, ensuring uninterrupted service necessitates shaping radio access capacity to adapt to the variable channel quality characterizing wireless links.

In this paper, we have presented a comprehensive modeling framework that considers both data and beamformer control information to accurately dimension the required rate on the fronthaul link. We have delved into techniques to adjust access parameters to align with the transport link capacity. Finally, we have validated these techniques using real data obtained from a D-band fronthaul link, demonstrating their efficacy in practical scenarios.

\section*{Acknowledgments}
\small{The research in this paper has been carried out in the framework of Huawei-Politecnico di Milano Joint Research Lab. The Authors acknowledge L. Resteghini, F. Capeletti, L. Luini and Huawei Milan research center for the collaboration.
The work of E. Moro and I. Filippini was partially supported by the European Union under the Italian National Recovery and Resilience Plan of NextGenerationEU, partnership on “Telecommunications of the Future” (PE00000001 - program “RESTART”, Structural Project 6GWINET).
}
\bibliographystyle{ieeetr}
\bibliography{Bibliography_short.bib}

\end{document}